# Information Technology Utilization in Environmentally Friendly Higher Education

Leon Andretti Abdillah[1*], Arif Ainur Rofiq[2], and Dian Eka Indriani[3]

[1]Department of Information Systems, Universitas Bina Darma, Jalan Ahmad Yani, Palembang, Indonesia
[2]Department of Islam Guidance and Counseling, UIN Sunan Ampel Surabaya, Jalan Ahmad Yani, Surabaya, Indonesia
[3] Pancasila and Civic Culture Education, STKIP PGRI Bangkalan, Jalan Soekarno Hatta, Bangkalan, Indonesia
[*]leon.abdillah@yahoo.com, rifainurrofiq@uinsby.ac.id, dianindriani79@gmail.com

Abstract: The awareness of an environmentally friendly learning process has been of concern lately. IT offers some applications that are able to provide better green education environment in higher education sectors. This research involves top social information technology applications, DropBox and WordPress. DropBox is the most popular cloud storage in the world. Meanwhile, WordPress is top content management systems at the moment. This research employs the utilization of those two applications in students assignments and presentation medium. The study observed 150 sophomore students in computer science faculty. The results show that, involving DropBox and WordPress in higher education learning activities able to significantly reduce paper and ink usage. The combination of those two top applications create a very handy and comfortable environment for students in higer education. This strategy reduce the consumption of papers and inks and are accepted well by most of the students.

## 1 INTRODUCTION

The development and progress of information technology (IT) can be used for various purposes. In the field of government, IT has been used to cut bureaucracy and complicated procedures. In the business field, IT has been widely used as a virtual media for sellers and buyers to make online transactions or online shopping. The trend of using information technology such as the internet has also experienced a significant increase in transportation sector as online ride sharing. Among those examples authors would like to define the term of "IT utilization" as involving IT applications in daily activities. In this article, we involving some of IT applications for education field.

According to Internet World Stats, Asia continent is the region that has the most internet users (Internet World Stats, 2018b) followed by Europe, Africa, Latin America/Carribean, North America, Middle East, and Oceania/Australia. In Asia there are 48.7% of poulation are internet users (Table 1) or almost a half the population.

Among those data, Indonesa rank number 5 after China, India, United States, and Brazil. Indonesian internet users per December 2017 (Table 2) are 143,260,000 users (Internet World Stats, 2018a).

Table 1: World Internet Usage and Population Statistics.

| World Regions | Internet Users | Internet Users % |
|---|---|---|
| Africa | 453,329,534 | 10.9 % |
| Asia | 2,023,630,194 | 48.7 % |
| Europe | 704,833,752 | 17.0 % |
| Latin America / Caribbean | 437,001,277 | 10.5 % |
| Middle East | 164,037,259 | 3.9 % |
| North America | 345,660,847 | 8.3 % |
| Oceania / Australia | 28,439,277 | 0.7 % |
| WORLD TOTAL | 4,156,932,140 | 100.0 % |

In the early 2018 period, it was known that internet users in all the world reached more than 4 (four) billion people (Kemp, 2018) or equal to 53% (more than half world population).

Data on the number of university students enrolled in Kemenristekdikti in 2017 totaled 6,924,511 students (Ristekdikti, 2018). Learning activities on the college campus that have been carried out conventionally involve the provision of teaching materials, daily examinations, midterms, semester exams, and work assignments (both individual and group assignments). With that number of students, you can imagine how much paper and ink are used for the printing needs of these student assignments.

Table 2: Top 20 Countries With Highest Number of Internet Users - December 31, 2017.

| No | Country or Region | Internet Users 31 Dec 2017 |
|---|---|---|
| 1 | China | 772,000,000 |
| 2 | India | 462,124,989 |
| 3 | United States | 312,322,257 |
| 4 | Brazil | 149,057,635 |
| **5** | **Indonesia** | **143,260,000** |
| 6 | Japan | 118,626,672 |
| 7 | Russia | 109,552,842 |
| 8 | Nigeria | 98,391,456 |
| 9 | Mexico | 85,000,000 |
| 10 | Bangladesh | 80,483,000 |
| 11 | Germany | 79,127,551 |
| 12 | Philippines | 67,000,000 |
| 13 | Vietnam | 64,000,000 |
| 14 | United Kingdom | 63,061,419 |
| 15 | France | 60,421,689 |
| 16 | Thailand | 57,000,000 |
| 17 | Iran | 56,700,000 |
| 18 | Turkey | 56,000,000 |
| 19 | Italy | 54,798,299 |
| 20 | Egypt | 48,211,493 |

Furthermore, people will continue the trend of reducing their need to rely on costly hardware and infrastructure by placing files and applications in the cloud (Drake, 2018). Cloud storage will reduce the cost for investment in hardware and infrastructure. Cloud computing is viewed as a method for achieving efficiency and cost savings through the use of commodity IT infrastructure (Waddington et al., 2013).

This paper will discuss the role of IT in supporting environmentaly friendly higher education. Environmentally friendly higher education is everyday behavior that is applied or carried out in a university that has a positive impact on the environment that does not damage the environment.

The involvement of popular social information technology in the lectures gained a very good response from the learners (Abdillah et al., 2018). Author involves 2 (two) famous applications in social information technology. Those applications are DropBox and WordPress.

One of the most popular cloud repository is DropBox. A research showed that DropBox is currently the most popular provider (Drago et al., 2012) of cloud-based storage systems. The second application is WordPress. WordPress is a free installation resource which has many useful plug-ins, comment spam-fighting features, and user-friendly interface (Hong, 2008). Furthermore, bloggers are able to customize WordPress template and script according to their interest.

The combination of those two top applications create a very handy and comfortable enrivnment for students in higher education. With the use of no cost, high level of popularity, and ease of use, the two top applications can represent the use of IT or "IT utilization", especially in higher education.

Some previous studies have been reported in the field of IT and higer education, such as: 1) Social social network in blended learning (de Jorge Moreno, 2012, Abdillah, 2016a), 2) Students learning center and course management system (Dougiamas and Taylor, 2003, Abdillah, 2013), 3) Managing information and knowledge sharing (Abdillah, 2014), 3) Enriching course materials by using innovative learning resource for college students (Burke and Snyder, 2008, Abdillah, 2017).

The next section of this paper is research methods (section II) followed by results and discussions (section III). The last section is conclussion as section IV.

## 2 RESEARCH METHODS

In research methods section, Authors present the college students participants, the course subects, and the learning activities involved in this study.

### 2.1 College Students Respondents

The respondents of this study were students in level 3 (three) or fifth and sixth semester students. Those students are students who have taken elective courses. The elective courses taken will be the basic material for student research activities which then become the main theme or topic of their thesis. Total students who were respondents were 150 students who took three courses.

Of a number of elective courses available, the subjects that get the highest grades will be prioritized as the student thesis research theme. If there are several elective courses that get the same grades, then students will consult more closely with their academic advisers.

## 2.2 Course Subjects

Author involved 3 (tree) subjects in infotmation systems study program. Those cource subjects are: 1) Customer Relationship Management (CRM), 2) Supply Chain Management (SCM), and 3) Systems Analysis and Design (SA&D).

All courses are taught for 16 (sixteen) meetings in about 4 (four) months. CRM and SCM courses have a weight of 2 (two) credits. While SA&D courses have a weight of 4 (four) credits.

The content of CRM (Anderson and Kerr, 2002, Lindstrand et al., 2006, Buttle, 2009, Abdillah, 2018a) are as follow: 1) Introduction, 2) Basic Concept of CRM, 3) The Customer Service/Sales Profile, 4) Managing Your Customer Service/Sales Profile, 5) Choosing Your CRM Strategy, 6) Managing and Sharing Customer Data, 7) Tools for Capturing Customer Information, 8) Service-Level Agreements, 9) E-Commerce: Customer Relationships on the Internet, 10) Managing Relationships Through Conflict, 11) Fighting Complacency: The "Seven-Year Itch" in Customer Relationships, 12) Resetting Your CRM Strategy, 13) Presentations, and 14) Final Exam.

SCM course consist of (Simchi-Levi et al., 1999, Firdaus et al., 2015, Abdillah, 2018b): 1) Introduction, 2) Basic Concepts, 3) The Role of Purchasing in an Organization, 4) Creating & Managing Supplier Relationships, 5) Strategic Sourcing For Successful SCM, 6) Demand Forecasting & Collaborative Planning, Forecasting, & Replenishment, 7) Inventory Management, 8) Transportation Management, 9) Vendor Management, 10) Warehouse Management, 11) Cross Docking, 12) Third Party Logistics (3PLs), 13) IT in Supply Chain, 14) Presentations, and 15) Final Exam.

While for SA&D (Whitten and Bentley, 2007, Kendall and Kendall, 2011, Dennis et al., 2012a, Dennis et al., 2012b, Abdillah, 2016b), the lessons are: 1) Part One Planning Phase (The Systems Analyst and Information Systems Development, Project Selection and Management), 2) Part Two Analysis Phase (Requirements Determination, Use Case Analysis, Process Modeling, Data Modeling), 3) Part Three Design Phase (Moving Into Design, Architecture Design, User Interface Design, Program Design, Data Storage Design), and 4) Part Four Implementation Phase (Moving Into Implementation, Transition To The New System, The Movement To Objects).

## 2.3 Learning Activities

This study applies a blended learning approach that combines the process of conventional learning (face to face) activities with electronic based learning.

Out of sixteen meetings, more than half were face-to-face meetings. There are 2 (two) meetings that are indeed allocated for e-learning activities. if there is a lecture schedule that exists during the national holiday period, then the schedule will be held with an e-learning approach.

In conventional learning, students will study together with students in certain classes (around 20-40 students). In each class, a number of groups will be created. Each group will consist of 4 (four) to 8 (eight) students who will discuss a particular theme. The theme will be given in the early conventional meeting by lecturer.

## 3 RESULTS AND DISCUSSIONS

The results and discussion section will explain a number of results obtained after the research is completed. This section consists of 1) Characteristics of student respondents, 2) Use of blogs, 3) Use of cloud repositories, 4) Evaluation of blog usage, and 5) Evaluation of cloud repository usage

## 3.1 College Students Characateristics

Among those 150 students, 47,33% of them are taking CRM course subject, 40% are taking SCM course subject, and the rest of them or equal to 12,7% are taking SA&D course subject.

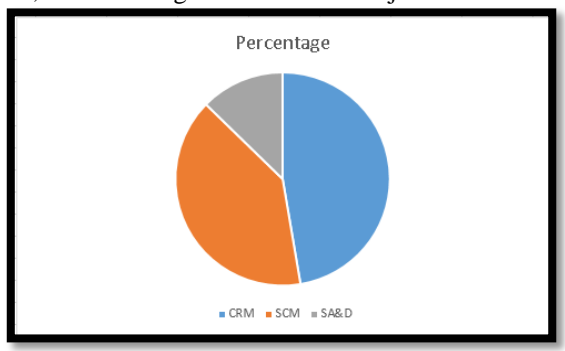

Figure 1: Lecturer Blog Page.

Meanwhile, most of the college participants were dominated by male students by 56% or 84 students,

while female students amounted to 44% or 66 students.

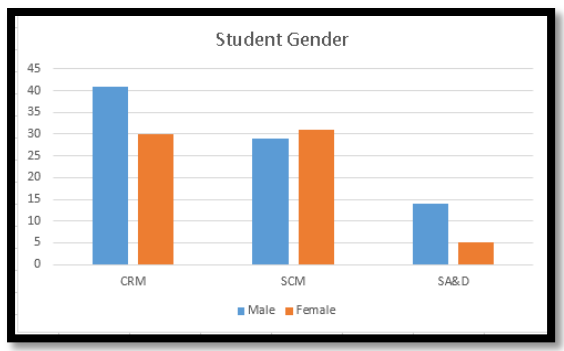

Figure 2: Lecturer Blog Page.

## 3.2 Blog

The use of blogs in learning provides alternatives that are cheap and fast. Blogs are generally offered free or without fees. Blogs have also been completed with templates that can be selected according to the needs and desires of the lecturers or students. More so blogs can be collaborated with both social media and the cloud repository.

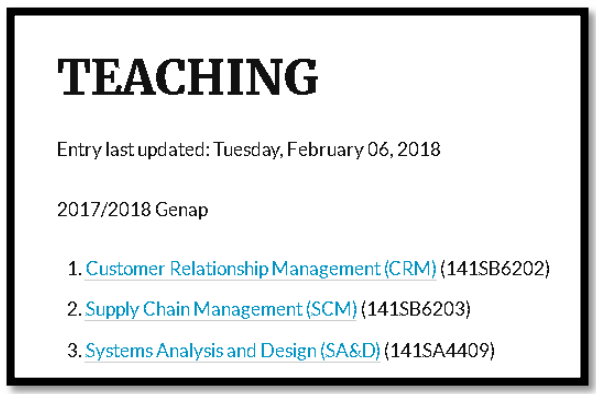

Figure 3: Lecturer Course List Blog Page.

Lecturers can utilize blog media for lecture materials, group lists, as well as a list of students who take certain courses. If needed, the lecturer can also add links that are considered useful or related to the course.

While students will work on weekly group assignments. Each task will be made into a post on the blog or in cloud storage. Every student who joined a flow group posted the assignment on their respective blogs.

## 3.3 Cloud Repository

During the lecture there were a number of assignments given to students. In the conventional lecture system these tasks will be collected in the form of volumes of papers. In this study, the authors utilize cloud storage media.

Placing our files in the cloud has many benefits (Mitroff, 2016). Not only for documents, but also photos, videos, and others kind of files. Almost popular cloud repositories at the moment provide some free storage, paying for extra space, and and the most important thing is the ability to synchronize between devices. Files stored in the cloud can be accessed anywhere and anytime as long as there is an internet connection.

In every week assignment, some of the assignments will be posted in the blog, and some others will be stored in the cloud repository and the URL will be linked into blog.

Cloud services provide 3 (three) main services (Fikri et al., 2015): 1) Software as a Service (SaaS), 2) Platform as a Service (PaaS), and 3) Infrastructure as a Service (IaaS). This study that is involves DropBox is close to SaaS.

## 3.4 Blog Usage Evaluation

During the exam the lecturer will provide a questionnaire regarding the involvement of blogs and cloud repositories.

Lecturers can use blogs as a medium to place lecture materials that will be given to students. Based on the questionnaires distributed, it is known that most of the students really like and like to get lecture materials through blogs (figure 4).

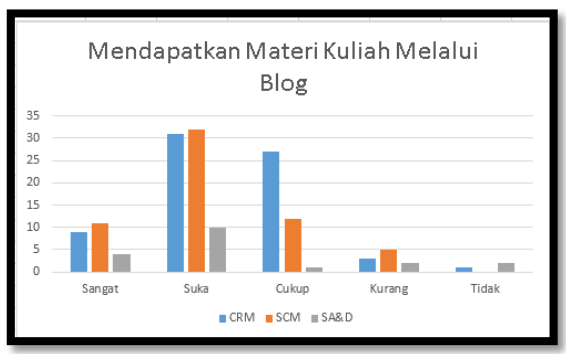

Figure 4: Student responses in getting lecture material through a blog.

Blogs can also be used to view the tasks of the group assignments. Each assignment will be grouped together and collected individually in the form of a blog post. Based on the questionnaires distributed, it is known that the majority of students are very fond and like to work on lecture assignments through blogs (figure 5).

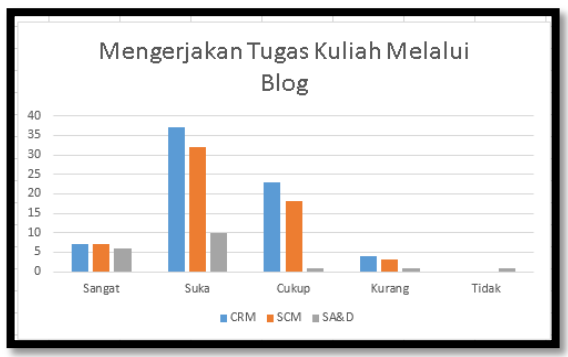

Figure 5: The response of students working on college assignments through blogs.

### 3.5 Evaluation DropBox Usage

Cloud repository is also used by students to store lecture tasks. A number of operating systems have provided each cloud storage service. Apple has an iOS platform that has iCloud, Google with Android has Google Drive, and Microsoft Windows provides OneDrive.

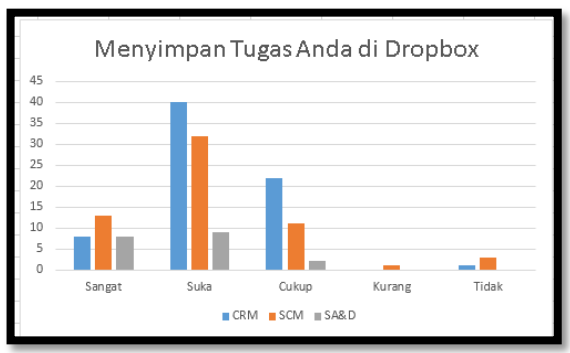

Figure 6: Students responses in storing their assignment in DropBox.

All cloud repositories must be accessed with an email affiliated with them. for this study, the authors prefer to use DropBox which can run on all platforms. DropBox provides huge storage media up to 10 (ten) TB for free.

Based on the questionnaires distributed, it is known that most of the students really like and like to keep their college assignments using DropBox (figure 6).

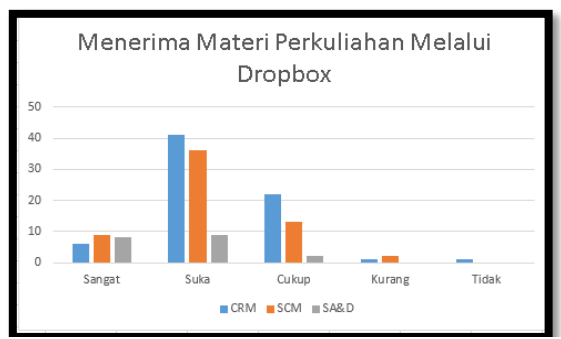

Figure 7: Students responses in getting their lecture materials by using DropBox.

In addition, it is known that most students also like and like to receive lecture material from lecturers through DropBox (figure 7).

## 4 CONCLUSIONS

Based on the results of the research that has been done, conclusions can be drawn, as follows: 1) The application used is an application that is free or does not require the cost of both software costs, hardware costs, and maintenance fees, 2) The use of WordPress and DropBox in the learning process greatly minimizes the use of paper and ink, and 3) Most students really like to receive lecture materials and work on their lecture tasks through Blog (WordPress) and Cloud Repository (DropBox).